\documentclass[aps,prb,reprint,superscriptaddress]{revtex4-1}

\usepackage{graphicx}% Include figure files
\usepackage{float}
\usepackage{gensymb}
\usepackage[mathscr]{euscript}
\usepackage{amsmath}

\begin{document}
\title{Observation of a superconducting glass state in granular superconducting diamond}

\author{G. M. Klemencic}
\email[KlemencicG@cardiff.ac.uk]{}
%\thanks{}
\affiliation{School of Physics and Astronomy, Cardiff University, Queen's Buildings, The Parade, Cardiff, CF24 3AA, UK}

\author{J. M. Fellows}
\affiliation{School of Physics, HH Wills Physics Laboratory, University of Bristol, Tyndall Avenue, Bristol, BS8 1TL, UK}

\author{J. M. Werrell}
\affiliation{School of Physics and Astronomy, Cardiff University, Queen's Buildings, The Parade, Cardiff, CF24 3AA, UK}

\author{S. Mandal}
\affiliation{School of Physics and Astronomy, Cardiff University, Queen's Buildings, The Parade, Cardiff, CF24 3AA, UK}

\author{S. R. Giblin}
\affiliation{School of Physics and Astronomy, Cardiff University, Queen's Buildings, The Parade, Cardiff, CF24 3AA, UK}

\author{R. A. Smith}
\affiliation{School of Physics and Astronomy, University of Birmingham, Edgbaston, Birmingham B15 2TT, UK}

\author{O. A. Williams}
\affiliation{School of Physics and Astronomy, Cardiff University, Queen's Buildings, The Parade, Cardiff, CF24 3AA, UK}

\date{\today}

\begin{abstract}
The magnetic field dependence of the superconductivity in nanocrystalline boron doped diamond thin films is reported. Evidence of a glass state in the phase diagram is presented, as demonstrated by electrical resistance and magnetic relaxation measurements. The position of the phase boundary in the  $H$-$T$ 
plane is determined from resistance data by detailed fitting to zero-dimensional fluctuation conductivity theory. This allows determination of
the boundary between resistive and non-resistive behavior to be made with greater precision than  the standard \textit{ad hoc} 
onset/midpoint/offset criterion.
\end{abstract}

\pacs{}
%\keywords{}

\maketitle

\section{Introduction}{\label{Introduction}}
%\subsection{}
%\subsubsection{}

Historically, the experimental observation of a proposed superconducting glass state was first made by M{\"u}ller, Takashige and Bednorz in powder samples of La$_{2}$CuO$_{4-y}$:Ba~\cite{muller1987flux}. The authors noted the existence of frustration in granular systems that gives rise to a ``superconductive glass'' that is equivalent to an XY spin-glass~\cite{ebner1985diamagnetic}, demonstrating the existence of degenerate states. In line with the essential features of the analogous spin-glass system, the authors reported a logarithmic decay of the remanent magnetization, and an irreversibility line -- the boundary in the $H$-$T$ plane between reversibility and
irreversibility in magnetization -- which obeyed $H \propto (1-T/T_c)^{3/2}$. This observed power law of the irreversibility line was thus likened to the de Almeida-Thouless line which separates the metastable and stable regions of the phase diagram in related models of spin-glasses~\cite{fischer1993spin}. 

Similar behavior regarding the irreversibility line and magnetic relaxation was subsequently observed in single crystal Y-Ba-Cu-O, and an alternative interpretation was proposed by Yeshurun and Malozemoff in the form of a thermally activated flux-creep model~\cite{yeshurun1988giant}. Tinkham then extended this work to propose a phenomenological model that includes the effects of pinning and reproduces the observed power 
law~\cite{tinkham1988creep}. Other theories purporting to explain the form of the irreversibility line in this kind of superconductor include vortex-lattice melting\cite{blatter1996low} and the existence of a disordered vortex glass\cite{fisher1989vortex, huse1992superconductors, lin1994vortex} state arising from random pinning sites.  Importantly, these competing theories share in common the idea of frustration of the superconducting order parameter, leading to a ubiquitous $3/2$ power law in the irreversibility line, so one may in general term this behavior ``glassy superconductivity'' without understanding the nature of the correlations specific to each.%dwelling too much on its particular microscopic origins.

Whilst the majority of reports of irreversible magnetic behavior have focused on high-$T_c$ compounds, Oppermann~\cite{oppermann1987quantum} has pointed out that the signatures of a glassy state should also be expected  in the vicinity of a metal-insulator transition,
regardless of the magnitude of $T_c$. Additionally, the existence of an irreversibility line has been studied in low-$T_c$ type II materials such as
Nb$_3$Sn and Nb-Ti~\cite{suenaga1991irreversibility}, Nb~\cite{schmidt1993applicability}, MgB$_2$~\cite{fuchs2001upper}, 
PbMo$_6$S$_8$ and Ba$_{0.25}$Pb$_{0.75}$BiO$_3$~\cite{zheng2003determination}.
The irreversible behavior in these metallic films has mainly been attributed to flux lattice melting. The boron-doped nanocrystalline diamond (BNCD) films studied here have a granular morphology~\cite{gajewski2009electronic, marevs2008grain, janssens2011separation, klemencic2017fluctuation} similar to sintered powders, and are close to a metal-insulator transition~\cite{takano2005superconductivity, klein2007metal, achatz2009low}, so that the observation of glassy superconductivity is perhaps not surprising. It is a novel observation for this material nonetheless, explicitly using the understanding of our previous work which used fluctuation spectroscopy to probe material properties that are governed by the morphological
granularity~\cite{klemencic2017fluctuation}. 

In this paper, we describe the measurement of the magnetic field-dependence of superconductivity in BNCD films, where the granularity may be controlled by the film thickness. We have found evidence in support of a superconducting glass in the phase diagram which we have confirmed with electrical resistivity and magnetic relaxation measurements. We observe a quasi de Almedia-Thouless-type line in $H_{c2}(T)$, and confirm the proposed glassy superconductivity by observing a logarithmic decay in the magnetization over time. Such glassy dynamics can be attributed to frustration in a system of weakly coupled superconducting clusters, each of which acts as an individual $U(1)\cong O(2)$ spin~\cite{ebner1985diamagnetic}.

\section{Material growth and experimental method}

The BNCD films, described in detail elsewhere~\cite{klemencic2017fluctuation}, are grown on SiO$_2$-buffered (100) silicon wafers by microwave plasma assisted chemical vapour deposition~\cite{williams2008growth}. Prior to growth, the substrates were seeded by ultrasonification in a monodisperse aqueous colloid of 5~nm diameter nanodiamond particles~\cite{williams2007enhanced}. During growth, the substrate temperature was $\sim$720$\degree$C in a plasma composed of 3\% methane in hydrogen, with trimethylboron as the source of boron (B/C ratio of 12800~ppm). The chamber pressure and microwave power were 40~Torr and 3.5~kW respectively. In this paper, we have focused our attention on the thickest sample, though other samples show qualitatively similar behavior. Scanning electron micrographs showed the thickness of this sample to be 564~nm, with a mean grain diameter of 102 $\pm$ 24~nm. 

All temperature dependent measurements relating to the magnetic field dependence of the superconductivity were made using a Quantum Design physical property measurement system with a base temperature of 1.9~K. Resistance measurements were performed using silver paste contacts made to the sample surface in a four-wire Van der Pauw configuration. Vibrating sample magnetometry was used to measure the magnetization. All measurements were made with the magnetic field applied perpendicularly to the samples.

\section{Results}

\subsection{Electrical resistance}

\begin{figure}
\includegraphics[width=8.6cm]{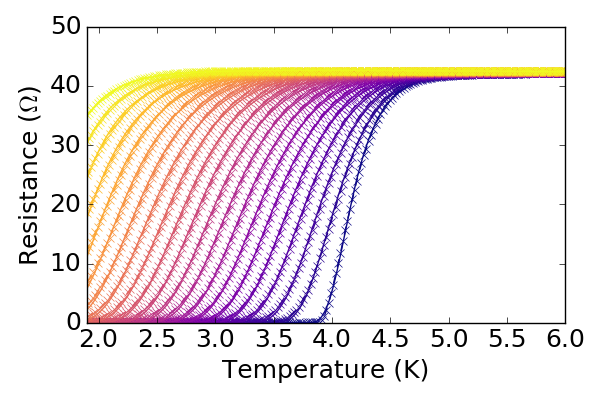}
\caption{\label{RTH564nm}Resistance as a function of temperature for applied magnetic fields in the range 0--4~T in 0.2~T increments. Fit lines (solid) through the data follow the analysis technique discussed in the text.} 
\end{figure}

\begin{figure}
\includegraphics[width=8.6cm]{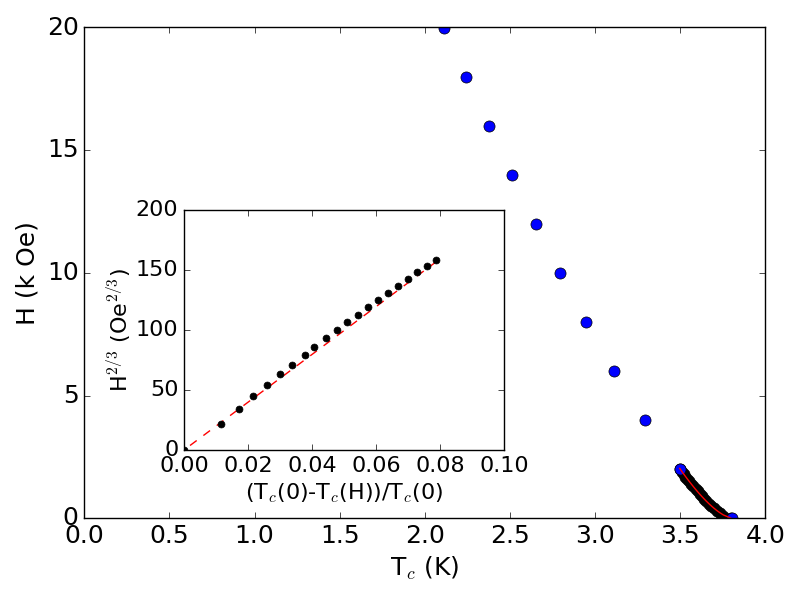}
\caption{\label{HT564nm}Magnetic field as a function of transition temperature for a 564~nm thick BNCD film. Inset: rescaled $H(T)$ clearly showing
the $H^{2/3}$ power law behavior.} 
\end{figure}

Fig.~\ref{RTH564nm} displays the resistance as a function of temperature measured in an applied magnetic field in the range $0$--$4$~T with field increments of 0.2~T. Closely spaced field increments of 0.02~T (not shown in Fig.~\ref{RTH564nm} for clarity, but present in Fig.~\ref{HT564nm}) were also used below 0.2~T. 
In order to make a more rigorous determination of $T_c$, we have fitted the fluctuation part of the conductivity to the quasi-zero dimensional scaling 
behavior, $\sigma_\text{fl}\sim(T-T_c)^{-3}$, discussed in detail elsewhere~\cite{klemencic2017fluctuation}. The critical temperatures so determined 
give a precise measure of the temperature at which the conductivity diverges for a given field. This is more accurate than determining
critical behavior through the more common onset-midpoint-offset (90\%-50\%-10\%) criterion~\cite{fuchs2001upper}. Indeed, each $R(T)$ trace in Fig.~\ref{RTH564nm} is displayed along with the
associated fit line, showing a remarkable agreement between the theoretical prediction and the experimental data. 

The resulting field induced transition temperature is shown in Fig.~\ref{HT564nm}. It has been shown that this can be interpeted as the irreversibility line, $H_\text{irr}(T)$~\cite{suenaga1991irreversibility, watanabe1992irreversibility, zheng2000irreversibility, zheng2003determination}; as the magnetization becomes reversible, a supercurrent is no longer supported, and a finite resistance is measured. There is a clear upturn in the 
$H_\text{irr}(T)$ curve, which closely follows the quasi de Almeida-Thouless $H^{2/3}$ power law, as shown more clearly in the inset. 

\subsection{Magnetic relaxation}

To confirm the observation of glassy superconductivity in this system, we have also measured the relaxation of the magnetization, and observed the logarithmic decay that is expected from the analogous frustrated model.

The procedure for measuring the magnetic relaxation described by Yeshurun, Malozemoff and Shaulov~\cite{yeshurun1996magnetic} was followed. The sample was cooled to the measurement temperature in zero applied field. A magnetic field smaller than the irreversibility field, but larger than the minimum sample magnetization, was then applied, followed by a step decrease that ensured a reversal of the flux profile. The values of these fields were determined by measurement of a magnetic hysteresis loop at the appropriate temperature. The magnetic moment was then monitored as a function of time by vibrating sample magnetometry, where the amplitude of sample translation was minimised to avoid possible field inhomogeneity.

\begin{figure}
\includegraphics[width=8.6cm]{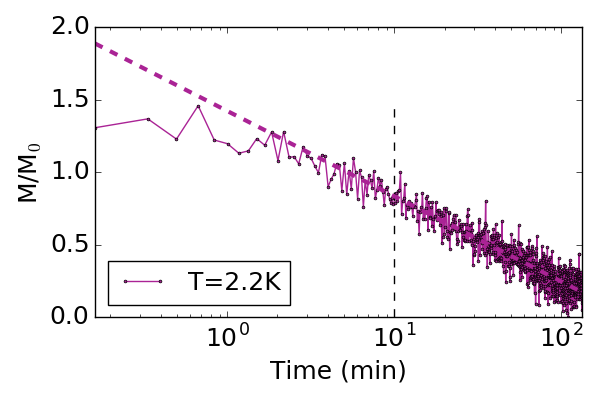}
\caption{\label{Relaxation564nm}Magnetic relaxation measured at 2.2~K for a 564~nm thick BNCD film. The vertical dotted line indicates that
fitting to a logarithmic decay is performed after a transient period of 10 minutes. Similar logarithmic decay is observed over a range of 
temperatures up to $T_c$.} 
\end{figure}

The normalised magnetic relaxation at 2.2~K is shown in Fig.~\ref{Relaxation564nm}. For this measurement, after cooling in zero field, an initial applied field of 96~Oe was followed by a step change to 33~Oe. The dashed line is a linear fit to the normalised magnetic moment as a function of $\ln t$. We observe a logarithmic decay of the magnetic moment as expected for the analogous frustrated model to which we liken the observations.

\begin{figure}
\includegraphics[width=8.6cm]{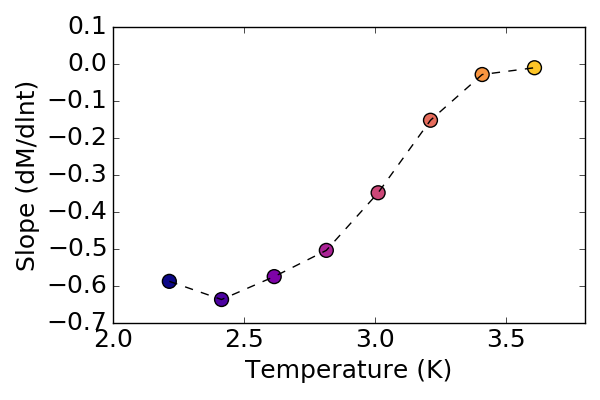}
\caption{\label{Slopes564nm}Gradient of the magnetic relaxation of a 564~nm thick BNCD film as a function of temperature.} 
\end{figure}

The slope of the magnetization $M(t)$ vs $\ln{t}$ is plotted as a function of temperature in Fig.~\ref{Slopes564nm}; we see that the relaxation rate increases as the superconducting transition is approached. The non-linearity of this curve means that the present data set cannot be directly compared to the Anderson-Kim model, in which one expects the magnetization $M(t)$ to be proportional to $1-\frac{k_\textsc{b}T}{U_0}\ln(t/t_0)$ \cite{andersonkim1964theory, yeshurun1996magnetic} i.e. the slope should be linear in temperature. Further investigation will be required to determine which
model is able to explain the glassy behavior of this system, and thus provide more insight into its microscopic origin.

\section{Discussion}

Generally, reports of the field dependence of the resistive transition in BNCD to date have shown that the transition width broadens with applied magnetic  field; the onset of the superconductivity has been shown to be less sensitive to the field than the tail in the resistance that grows with increasing field~\cite{takano2004superconductivity, zhang2014global, wang2006superconductivity}. Scanning tunnelling microscopy of CVD-grown BNCD films well below $T_c$ has shown that the temperature dependence of the energy gap in polycrystalline~\cite{dahlem2010spatially} and epitaxial~\cite{sacepe2006tunneling} boron doped diamond samples both broadly agree with BCS-like superconductivity. In BNCD films, a strong correlation between the local superconductivity and the granular structure has been shown, with grain boundaries perpendicular to the direction of film growth~\cite{dahlem2010spatially}. The energy gap showed a smooth variation across a grain, which is believed to be indicative of slowly varying doping concentration. The grain boundaries, however, are reported as having predominantly metallic-like conductance. It is therefore appropriate to consider our BNCD sample as a disordered collection of coupled superconducting grains. We thus interpret the results presented here as likely being due to a superconducting glass arising from frustration in a model of weakly linked superconducting clusters, as described by Ebner and Stroud~\cite{ebner1985diamagnetic}. 

The samples studied here are not unique in terms of the existence of a quasi de Almeida-Thouless line in the $H(T)$ diagram; there is some evidence in the literature that other BNCD samples show similar behavior under the influence of a magnetic field~\cite{takano2004superconductivity, takano2005superconductivity, sidorov2005superconductivity, zhang2014global}, but this feature has never before been commented upon.  While discussion of, and experiments pertaining to, glassy states in superconductors are common for high-$T_c$ compounds, they are also expected to be observable in granular materials, or materials near the metal-insulator transition~\cite{oppermann1987quantum}. The BNCD samples studied here fall into both of these categories, and so it is perhaps unsurprising that this glassy behavior manifests itself here, but this has not previously been recognised.

As an important side-note, it is common practice throughout the literature to calculate the coherence length in films such as these by comparing the $H(T)$ curve to the Werthamer-Helfand-Hohenberg (WHH) theory~\cite{helfand1964whh} of the upper critical 
field~\cite{ekimov2004superconductivity, zhang2014global} . According to WHH theory, the $H(T)$ curve should be linear near to $T_c$, which is evidently not the case in samples displaying this glassy behavior due to the curvature of the quasi de Almeida-Thouless line. It follows that fitting to WHH may not be a reliable way to determine the coherence length, and studies doing this may need to be reconsidered.

Finally, the signature of glassy superconductivity observed here has potential consequences for the use of BNCD in low temperature device engineering. The high Young's modulus of diamond -- even in the form of nanocrystalline films~\cite{williams2010high} -- makes it a good candidate for the fabrication of high frequency nanoelectromechanical resonators with which to study the fundamentals of quantum mechanics in a macroscopic object~\cite{etaki2008motion, o2010quantum}. Recently, an anomalous temperature dependence of the dissipation in a superconducting nanoresonator was observed, and was suggested to arise from the dynamical vortex behavior below $T_c$~\cite{lulla2013evidence}. It is possible, therefore, that careful measurement of the temperature-dependent dissipation of a glassy superconducting nanoresonator~\cite{bautze2014superconducting} would show complementary behavior to that observed in this study.

\section{Conclusions}

We have examined the magnetic field dependence of the superconductivity in BNCD. We extract values of the transition temperature in increasing magnetic fields, and clearly observe the behavior $H \propto (1-T/T_c)^{3/2}$. This observed power law, coupled with a logarithmic decay of the remanent magnetization, is attributed to a superconducting glass state resulting from the morphological granularity of our BNCD samples.

\begin{acknowledgments}
The authors gratefully acknowledge support by the European Research Council under the EU Consolidator Grant `SUPERNEMS'.
\end{acknowledgments}

\bibliography{Klemencic_2018}

\end{document}